\documentclass[fleqn,10pt]{wlscirep}

\PassOptionsToPackage{hyphens}{url}
\usepackage{graphicx}%
\usepackage{multirow}%
\usepackage{amsmath,amssymb,amsfonts}%
\usepackage{amsthm}%
\usepackage{mathrsfs}%
\usepackage[title]{appendix}%
\usepackage{xcolor}%
\usepackage{textcomp}%
\usepackage{manyfoot}%
\usepackage{booktabs}%
\usepackage{algorithm}%
\usepackage{algorithmicx}%
\usepackage{algpseudocode}%
\usepackage{listings}%
\usepackage[most]{tcolorbox}
\usepackage[
    type={CC},
    modifier={by},
    version={4.0},
]{doclicense}

\begin{document}

\title{Fixation-related potentials reveal that confusing program code elicits a late frontal positivity}

\author[1,+]{Annabelle Bergum}
\author[1,+*]{Anna-Maria Maurer}
\author[1]{Norman Peitek}
\author[2]{Regine Bader}
\author[2]{Axel Mecklinger}
\author[1,3]{Vera Demberg}
\author[4]{Janet Siegmund}
\author[1]{Sven Apel}

\affil[1]{Computer Science, Saarland University}
\affil[2]{Psychology, Saarland University}
\affil[3]{Language Science and Technology, Saarland University}
\affil[4]{Computer Science, University of Technology Chemnitz}

\affil[*]{maureran@cs.uni-saarland.de}
\affil[+]{These authors contributed equally to this work.}

\begin{abstract}
As software pervades more and more areas of our professional and personal lives, there is an ever-increasing need to maintain software and for programmers to efficiently write and understand program code. In the first study of its kind, we analyze \emph{fixation-related potentials} (FRPs) to explore the online processing of program code patterns that are confusing to programmers, but not to the computer (so-called \emph{atoms of confusion}), and their underlying neurocognitive mechanisms in an ecologically valid setting. Relative to clean counterparts in program code without an atom of confusion, confusing code elicits a late frontal positivity of about 400 to 700 ms after first looking at the atom of confusion. This frontal positivity resembles an event-related potential (ERP) component found during natural language processing that is elicited by unexpected but plausible words in sentence context. Thus, we suggest that the brain engages similar neurocognitive mechanisms in response to unexpected and informative inputs in program code and in natural language. In both domains, these inputs update a comprehender's situation model, which is essential for information extraction from a quickly unfolding input. Our results have far-reaching implications for programming and pave the way for interdisciplinary collaborations between software engineering and psycholinguistics.
\end{abstract}


\flushbottom
\maketitle

\begin{figure}[b]
    \doclicenseThis
\end{figure}

\section{Introduction}\label{sec:introduction}
Software forms an integral aspect of our everyday life. Thus, it is highly relevant to understand the processes that underlie skilled programming, including the creation and maintenance of software. One central aspect of programming is reading and comprehending program code, which is crucial for various programming activities, such as implementing new functionalities or fixing program errors~\cite{LaToza2006, Tiarks2011, Minelli2015, meyer2020detecting, meyer2019today}. During these activities, misconceptions about the code, such as confusion regarding its execution, purpose, or intended message, can lead to software failures with potentially fatal consequences~\cite{meyer2020detecting, meyer2017work}.

The main goal of the present study is to explore the processing of confusing elements in program code and their underlying neurocognitive mechanisms in an ecologically valid setting. Recently, various studies have identified a set of small, recurring patterns~\cite{Gopstein2017} prevalent in program code~\cite{Gopstein2018, medeiros2019investigation, pinheiro2023they}, the so-called \emph{atoms of confusion}. An \emph{atom of confusion} is a syntactically correct code pattern that is unambiguous for the computer to execute, but, due to its unintuitive structure, is confusing to programmers, who encounter difficulties in understanding the code and might even misinterpret it~\cite{Gopstein2017, Gopstein2020}.

\lstdefinestyle{mystyle}{
    backgroundcolor=\color{white},   
    commentstyle=\color{codegreen},
    keywordstyle=\bfseries\color{cyan},
    numberfirstline=true,
    numberstyle=\scriptsize\ttfamily\color{gray},
    stringstyle=\color{codepurple},
    basicstyle=\ttfamily\small,
    breakatwhitespace=false,         
    breaklines=true,                 
    captionpos=b,                    
    keepspaces=true,                 
    numbers=left,                    
    numbersep=5pt,                  
    showspaces=false,                
    showstringspaces=false,
    showtabs=false,                  
    tabsize=2,
    language=Java,
    captionpos=b}

\lstset{style=mystyle}
\begin{figure}[htb]
\hspace{15pt}
\begin{minipage}[t]{0.47\textwidth}

\begin{lstlisting}[caption={Confusing code containing an atom of confusion (\emph{indentation of the else block makes it difficult to match it to the correct if statement})},label={lst:atom_ambiguous},showlines=true]
boolean V1 = true;
boolean V2 = false;
int R = 3;
if (V1)
   if (V2)
      R = R + 2;
else
   R = R + 4;
\end{lstlisting}
\end{minipage}
\hfill
\begin{minipage}[t]{0.47\textwidth}  
\begin{lstlisting}[caption={Equivalent clean code without an atom of confusion (\emph{else block is unambiguously indented to make it easy to match it to the second if statement})},label={lst:atom_unambiguous}]
boolean V1 = true;
boolean V2 = false;
int R = 3;
if (V1)
   if (V2)
      R = R + 2;
   else
      R = R + 4;
\end{lstlisting}
\end{minipage}
\caption{A pair of corresponding code snippets with one containing an atom of confusion (left) and the other containing functionally equivalent code that is easy to understand (right).}\label{fig:atom}%
\end{figure}

In Figure~\ref{fig:atom}, we contrast an illustrative example of a confusing code snippet (containing an atom of confusion) with a clean code snippet (containing functionally-equivalent code without the atom of confusion). The atom of confusion is located in Lines~7 and 8 of the confusing snippet displayed in Listing~\ref{lst:atom_ambiguous}, in which the level of indentation of the \verb|else| block subtly influences human understanding. The difficulty that programmers may encounter is that the indentation suggests the \verb|else| block closes the first \verb|if| statement, which leads to the incorrect result of 
\verb|R = 7|. In the clean counterpart shown in Listing~\ref{lst:atom_unambiguous}, the indentation aligns with the functionally correct second \verb|if| statement. This eliminates the risk of confusion regarding the matching \verb|if| statement, and the output of the variable 
\verb|R| can be straightforwardly determined as \verb|3|, as several studies on atoms of confusion confirmed~\cite{olivera2020, Gopstein2017, Langhout2021}. It is important to note that the confusing and clean snippet conditions are both syntactically correct and semantically equivalent, but they differ in their risk that a human programmer may misunderstand their semantics due to the atom of confusion~\cite{Gopstein2017}. While such subtle differences may seem insignificant at first glance, they have caused severe bugs in the past, such as a mismatched indentation in Apple's iOS operating system that rendered its secure network connections vulnerable~\cite{Wheeler2014apple, Boyes2014}.
In general, several behavioral studies across different programming languages have shown that atoms of confusion require more time to process~\cite{olivera2020, Gopstein2017}, are more effortful to comprehend~\cite{Yeh2017, daCosta2023}, and lead to more errors~\cite{Gopstein2017, Yeh2022,Langhout2021}.

Despite the insights into atoms of confusion obtained from behavioral experiments, the underlying mechanisms of the phenomenon remain unclear. One missing element is our limited understanding of which neurocognitive mechanisms are engaged in response to the confusing program code that is causing the difficulty. Identifying these mechanisms would allow us to gain a deeper understanding of the principles behind misconceptions in program code, with far-reaching implications for programming practice. Specifically, this line of research can provide empirical evidence for refining coding conventions, programming language design, and educational practices, ultimately leading to more secure and reliable software.

The premise of our work is that our understanding of how humans process natural language could be the key to interpreting the processing of atoms of confusion and discovering the underlying neurocognitive mechanisms. Indeed, parallels between programming languages and natural languages have been observed in several studies, hinting at an overlap in activated brain areas between natural language (e.g., Brodmann Areas 21, 44, and 47) and program comprehension~\cite{Siegmund2014fMRI,Peitek2018:TSE, Duraes2016, Castelhano2019, Floyd2017}. In general, considering analogies with what is already known about human language processing and the processing of program code might prove to be a very fruitful avenue. 

The characteristics of program comprehension suggest that \emph{event-related potentials} (ERPs) might be particularly well suited to study atoms of confusion. ERPs are online electrophysiological responses to linguistic or other eliciting events. They can track cognitive processes even when unfolding quickly, which is the case for both reading program code and natural language text~\cite{Luck2014}. ERPs can provide a particularly informative way of exploring which neurocognitive processes underlie the processing of atoms of confusion, as there exists a rich literature in psycholinguistics that has related several ERP components to different cognitive processes that occur during language processing~\cite{swaab2012language}.

At present, it is unclear which aspect of the atoms of confusion causes the processing difficulties that have been observed for these elements. Is the difficulty (1) semantic in nature (i.e., the critical expression is unexpected and difficult to integrate with the meaning of the previous context, as in the sentence \textit{I take coffee with cream and dog})? In this case, the atom of confusion should elicit an enlarged N400 (i.e., a negative peak in the ERP approximately $400~\text{ms}$ after the eliciting event, similar to the N400 for the word \textit{dog} in the aforementioned example~\cite{Thornhill2012, Kuperberg2020, kutas1980reading, kasparian2016confusing}). 

Or should we rather think of atoms of confusion in terms of (2a) syntactic reanalysis similar to what is typically observed in garden path sentences, such as \textit{The horse raced past the barn fell}, which, like an atom of confusion, in fact only has a single correct interpretation that is, however, difficult to work out for a human, as there might be uncertainty about the function of the verb \textit{raced}. It can either be analyzed as a main verb of the sentence or alternatively as a past participle of a reduced relative clause \textit{The horse (that was) raced past the barn fell} (which is similar to indentation in our code example, where there is uncertainty about the program execution). In this case, we should expect to see a late posterior positivity (P600)~\cite{nunez2004, osterhout1992event}. Besides garden path sentences that involve syntactic reanalysis, the P600 is generally observed in sentences with syntactic errors, such as \textit{The child throw the toy}, but also in sentences with syntactic ambiguities, as in \textit{The woman persuaded to answer the door} relative to the syntactically unambiguous sentence \textit{The woman struggled to prepare the meal}~\cite{osterhout1992event, friederici2004relative}. In these cases, the P600 is typically interpreted as a correlate of a syntactic revision (e.g., passivizing the verb \textit{persuaded} and attaching it to a reduced relative clause), but it can also be evoked by semantic violations (see Paczynski and Kuperberg~\cite{paczynski2012multiple} for an example).

As a third alternative, one can think of atoms of confusion in terms of (2b) unexpected but plausible material, similar to materials in natural language that have been found to elicit a late frontal positivity in strongly constraining sentence contexts~\cite{Thornhill2012, brothers2020going}. In the sentence \textit{he bought her a pearl necklace for her collection}, the word \textit{collection} elicits a late frontal positivity, which is assumed to reflect the inhibition of the highly expected word \textit{birthday} and the integration of the unexpected word \textit{collection} in the current situation model -- an overall meaning representation that is constructed during language comprehension~\cite{kuperberg2016separate}. As outlined by Van Petten and Luka~\cite{van2012prediction}, both late components, the P600 and the late frontal positivity, are associated with unexpected linguistic inputs, but the main factor that distinguishes them is the plausibility of the input: The P600 is produced by highly implausible and anomalous words, whereas the frontal positivity is elicited by unexpected but plausible input.

Currently, it is unknown whether the confusion induced by atoms of confusion, as perceived by programmers, results (1) from the semantic integration of unexpected and challenging content (in which case we would expect to see an N400), (2) from syntactic ambiguity (see below), or (3) from something else entirely (producing an unexpected ERP signal). As for syntactic ambiguity, the confusion may arise from (2a) syntactic reanalysis (in which case we would expect to see a late posterior positivity) or from (2b) the integration of unexpected but plausible content into the situation model of the preceding context (in which case we would expect to see a late frontal positivity).Thus, we pose the following research question: 
\newtcolorbox{researchquestion}[1][]{%
  colback=gray!6,
  grow to right by=0mm,
  grow to left by=0mm,
  boxrule=0pt,
  boxsep=0pt,
  arc=0pt,
  breakable,
  enhanced jigsaw,
     borderline={0pt}{0pt}{black},
  #1,
}
\begin{researchquestion}
      Which ERP components identified in psycholinguistic research are elicited when processing atoms of confusion?
\end{researchquestion}

The premise of our work is that building on natural-language studies would help us understand the sources and processing of such confusion, illuminating the intricacies of program comprehension. This, in turn, would be invaluable in better supporting programmers through dedicated programming languages and curricula. For example, if atoms of confusion evoke a perceived syntactic ambiguity, this could be avoided by better design choices made by programming language designers (e.g., by requiring explicit brackets around the bodies of \verb|for|, \verb|while|, and \verb|if| statements); if they rather cause problems with semantic integration, this can be captured in dedicated training procedures within programming curricula and coding guidelines (e.g., defining clear and consistent rules that instruct programmers on the use of patterns). If we observe an entirely unexpected ERP signal, this could indicate that the confusion arising from the presence of an atom of confusion has no equivalent in natural-language comprehension, or that there is no unified cause for atoms of confusion. Thus, independent of the concrete results, it is highly relevant for the field of program comprehension to understand the electrophysiological signals and hence the underlying mechanisms of processing atoms of confusion. Looking at a broader picture, if we observe more and more analogies between processing program code and natural language text, this suggests that our premise of a methodological transfer between the two disciplines is worthwhile.

Since it is unclear which ERP component relates to processing confusing code, we employ a data-driven approach to detect differences between confusing and corresponding clean code without targeting a hypothesized component, as research on program comprehension and its underlying neurocognitive mechanisms is still in its infancy. Contrary to research on natural language comprehension, which builds on several decades of behavioral, neurophysiological, and brain imaging studies that have unraveled a variety of processing aspects and representational structures~\cite{kuperberg2019multimodal}, studying program comprehension faces unique challenges, such as a non-linear reading flow (e.g., program statements are processed not in lexical order but along the control and data flow of possible program executions)~\cite{Peitek2020, Aljehane2021, Busjahn2015, peitek2022correlates}. So far, only one study on program comprehension has investigated ERP responses to syntactic and semantic errors in program code: Using a token-wise presentation of code, Kuo and Prat~\cite{Kuo2024} identified P600 and N400 components for syntactically and semantically invalid program code, respectively. However, this setup of a token-wise presentation, while usually applied in reading studies with natural language materials~\cite{wen2021fast}, is too artificial and highly problematic for the analysis of program code: First, it enforces an unnatural linear reading flow. Second, with a token-wise presentation mode, each token is presented with a uniform duration, which allows the programmer only a cursory glance at the code but does not account for calculations that can strongly vary in complexity between tokens within a code snippet. Both aspects strongly compromise the construct and ecological validity of neurocognitive studies on program comprehension.

\begin{figure}[ht!]
\centering
\includegraphics[width=0.95\textwidth]{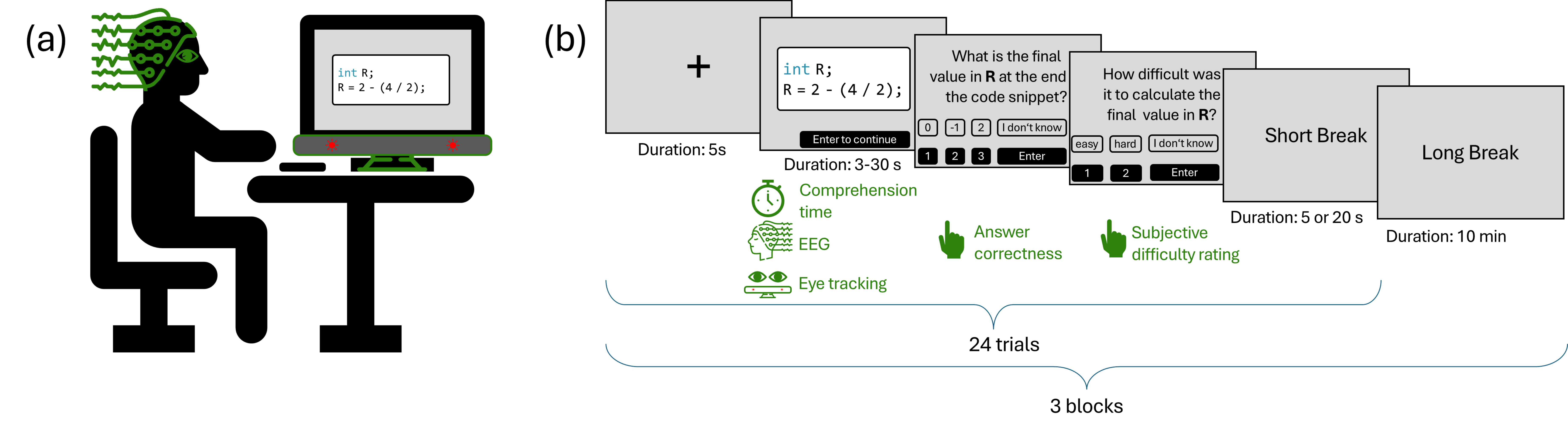}
\caption{
\textbf{Overview of the experiment setup and design} (EEG Icon by Freepik).
Part~\textsf{(a)} displays a symbolic image with the experiment setup of a participant with keyboard and EEG cap in front of an eye tracker attached to the bottom of the screen display.
Part~\textsf{(b)} visualizes the experiment design. The experiment included three blocks with $24$ trials each: fixation cross ($5~\text{s}$ to calibrate); snippet presentation ($3$--$30~\text{s}$, terminated by participant; the duration of snippet presentation is measured as comprehension time); answer correctness (the participant has to submit the output, i.e., the final value in variable \lstinline{R}); subjective difficulty rating (the participant has to indicate their perceived difficulty of performing the task); short break ($5$ or $20~\text{s}$) to reduce fatigue effects. After the block is finished, there is a long break of $10~\text{min}$ to reduce learning and fatigue effects.}
\label{fig:design}
\end{figure}

In an effort to increase the ecological validity of our study with regard to code presentation and code reading, we seek to present program code as a whole and allowed participants to read the code at their own pace. A major challenge in presenting a snippet of code all at once, however, is that the exact point in time at which the critical segment containing the atom of confusion is processed varies both between items and between participants and therefore cannot be assessed with conventional ERP averaging methods. That is, simply synchronizing electroencephalographical (EEG) activity with the onset of the presentation of the code snippet leads to a temporal misalignment of the ERP activity related to the processing of the atom within the trial period and, thus, diffuses the signal. This severely reduces statistical power and interpretability. 
To address this key challenge, we employ a more sophisticated and powerful approach based on \emph{fixation-related potentials} (FRPs). In the FRP approach, event synchronization is not realized by external stimuli~\cite{hutzler2007welcome}. Rather, eye movements and EEG are recorded simultaneously while the program code is presented as a whole, as illustrated in Figure~\ref{fig:design}a. The eye-tracking data are used to assess the participants' first eye fixations in the area of the atom of confusion (which is well-defined for every atom of confusion). This first fixation is then used for aligning the EEG signals. This way, we are able to determine the exact point in time at which information from the atoms is processed and can use it for synchronizing the electrophysiological data.

To uncover which neurocognitive mechanisms are engaged in response to processing confusing program code, we conducted an FRP experiment involving $24$ participants. We carefully adjusted the existing stimuli from previous work on atoms of confusion in code written in the Java programming language~\cite{Langhout2021} to optimize them for an FRP setup. In particular, each snippet was presented in a condition with and without an atom of confusion (across participants). Each pair of the snippet condition had a comparable number and length of code lines. Furthermore, to select suitable fixations for FRP analysis, we ensured that each snippet contained a single, clearly defined area of interest that included either the atom of confusion (\emph{confusing snippets}) or an equivalent without an atom (\emph{clean snippets}) (see Section~\ref{sec:methods} for details). The experiment consisted of three blocks, as illustrated in Figure~\ref{fig:design}b, in which we presented $24$ snippets (half confusing and half clean). We asked the participants to comprehend the snippets as quickly and accurately as possible in order to report the output. During the experiment, we collected comprehension time, answer correctness, and subjective difficulty ratings to verify that our confusing snippets were generally more difficult to process on the behavioral level than clean snippets. As illustrated in Figure~\ref{fig:design}a, we collected EEG and eye-tracking data. We used this data to calculate FRPs by mapping fixations to the areas of interest within the code snippets. We took special care to identify and correct eye movements and muscular artifacts by means of independent component analysis (see Section~\ref{sec:methods}, for further details). We synchronized the fixations with the EEG data stream to calculate FRPs. On the obtained FRPs, we tested for significant amplitude differences between confusing and clean code using pairwise cluster permutation tests. 

\section{Results}\label{sec:results}
We report on the first-ever FRP analysis of program comprehension by measuring and pinpointing the processing of confusing elements and their neural underpinnings during program comprehension in an ecologically valid setting. Our results show that confusing code elicits a larger frontal positivity between $390$ and $660~\text{ms}$ after first fixating the relevant area of the atom of confusion compared to fixating clean code (without an atom). We relate the identified component to the late frontal positivity occurring in natural language processing for unexpected but plausible input, which elicits an update of the situation model. 

\subsection{Behavioral results}
The behavioral data comprised comprehension time, answer correctness, and subjective difficulty ratings, where the last two were mapped to a binary categorical scale (the submitted answer was correct or the submitted rating was easy). The descriptive trial statistics are illustrated in the left part of Table~\ref{tab:lmer_results}.

In line with prior work, our behavioral results based on general linear mixed-effects regression (GLMER) models show that the confusing snippets took longer to comprehend, had fewer correct answers, and were rated less frequently as easy to understand, as displayed in the right part of Table~\ref{tab:lmer_results}. The full analysis results, including all fixed and random effects, are provided in the supplementary information, Appendix C.

\begin{table}[ht]
\caption{Overview of the behavioral data and results of general linear mixed-effect regression (GLMER) models: The first two columns show the descriptive statistics on trial-level, where answer correctness and subjective difficulty rating were mapped to a binary scale. The other columns present an overview of the regression model results. For comprehension time, an LMER with Gaussian error probabilities was applied on logarithmic data; for correct answers and subjective difficulty rating, we applied a binomial GLMER. All models used random intercepts and slopes for condition under participants and snippet number. They also include block number and the item order within the block as fixed predictors. The full regression model results can be found in supplementary information, Appendix C. One of 1,728 trials was excluded due to hardware malfunction.}\label{tab:lmer_results}%
\centering
\begin{tabular}{@{}llllll@{}}
	\toprule
	                             & \multicolumn{2}{l}{Trial statistics}   & \multicolumn{3}{c}{GLMER results for condition = confusing} \\ \midrule
	Trial count                  & Clean & $864$                   & Estimate           & Statistic     & Probability           \\
	                             & Confusing   & $863$                   & $\beta~\pm \mathit{SD}$     & $t$/$z$ value & $p$ value             \\ \midrule
	Comprehension time           & Clean & $11.2~\pm 7.4~\text{s}$ &                             &               &                       \\
	                             & Confusing   & $12.2~\pm 7.9~\text{s}$ & $~~\, 0.076~\pm 0.023$  & $t = ~~\,3.27$  & $~~~\,0.001$\\
	Answer correctness           & Clean & $88\% $                 &                    &               &                       \\ 
	                             & Confusing   & $70\% $               & $-1.572~\pm0.370$   & $z = -4.257$  & $<0.001$              \\
	Subjective difficulty        & Clean & $86\% $                 &                    &               &                       \\
	    rating                & Confusing   & $82\% $                 & $-0.789~\pm 0.398$ & $z = -1.983$  & $~~~\,0.047$              \\
	\bottomrule                     &              &                         &                    &
\end{tabular}
\end{table}

\subsection{Fixation-related potentials analysis}
In a first step, we performed an ERP analysis of the EEG time-locked to the onset of the snippets with the goal of testing whether, without the eye-tracking information, ERPs differ between confusing and clean code. Consistent with our assumption that ERPs cannot reveal meaningful differences when presenting a code snippet as a whole, we found no significant ERP differences between the two conditions (cf. supplementary information, Appendix B). 

\begin{figure}[ht!]
\centering
\includegraphics[width=0.95\textwidth]{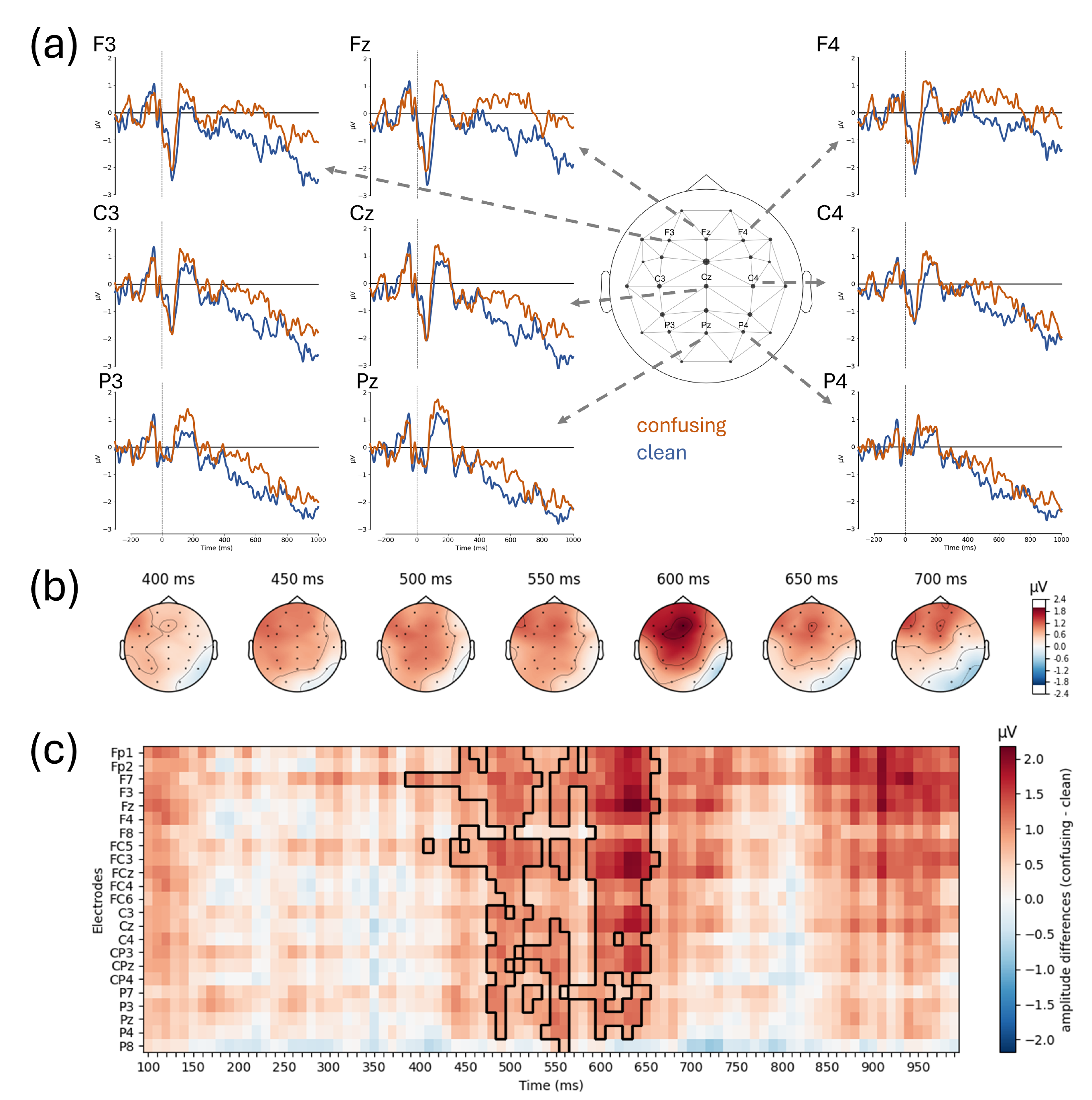}
\caption{\textbf{Results of the FRP analysis.} Part~\textsf{(a)} shows the FRP elicited by confusing (orange) and clean (blue) program code at nine scalp electrodes. The zero time points denote the onset of the fixation and positive voltages are plotted upwards. Part~\textsf{(b)} portrays the topographic distribution of the amplitude difference between confusing and clean code in consecutive $50~\text{ms}$ time intervals. The onset of the time interval is indicated above each map. As apparent from the figure, the frontal slow wave is most pronounced at frontal and left frontal recording sites. Part~\textsf{(c)} shows the amplitude differences between confusing and clean program code for all electrodes under investigation and all time points between $100$ and $1,000~\text{ms}$ after fixation onset. The significant cluster is surrounded by black lines. Note that the late amplitude differences at frontal recordings did not reach the cluster-based significance threshold.}
\label{fig:frp}
\end{figure}

Our primary goal was to determine the neurocognitive processes triggered by the presence of atoms of confusion in program code and to relate those processes to ERP components typically identified during natural language comprehension. In fact, there is a large delay and large variance in the point in time when participants first fixate on the critical region after the onset of the snippet: the first fixation on the critical region happens, on average, $3,772~\pm3,367~\text{ms}$ after stimulus onset in the clean condition, and $3,246~\pm3,084~\text{ms}$ in the confusing condition. The stimulus-onset-aligned ERPs, hence, cannot capture the first point in time at which information from the critical region is presumably processed. Fixation data show that on the first pass, participants fixated the critical region for $649~\pm594~\text{ms}$ in the clean condition and $805~\pm943~\text{ms}$ in the confusing condition. Therefore, we compared FRPs time-locked to the first fixation in the area of the atom of confusion and its counterpart without atom, spanning an epoch from $300~\text{ms}$ preceding the start of the fixation until $1,000~\text{ms}$ thereafter. For some trials, this epoch could not be extracted due to either low quality in the fixation data, no fixation in the area of interest, artifacts in the EEG signal, or a too short time period until the end of the snippet viewing phase. We excluded these trials from the subsequent analysis ($341$ of $1,728$). Then, we calculated the participants' FRP averages for both conditions across all trials. The grand average FRP data across all participants for $9$ electrodes at frontal, central, and parietal recording sites are presented in Figure~\ref{fig:frp}a.

As illustrated in Figure~\ref{fig:frp}, confusing code elicits a sustained frontal positivity relative to clean code. This positivity was most pronounced between $390$ and $660~\text{ms}$ at frontal recording sites (i.e., F3, Fz, F4). At posterior recording sites, the positive amplitude difference between confusing and clean code was attenuated in this time interval. Figure~\ref{fig:frp}b shows the difference in amplitude for all electrodes in the form of topographical distributions in consecutive $50~\text{ms}$ time intervals between $400$ and $700~\text{ms}$. Positive differences between both conditions were also present beyond $800~\text{ms}$ after fixation onset at some left frontal recording sites (e.g., F7 and F3), though they were less consistent. We statistically analyzed these effects using a cluster-based paired permutation t-test, which identifies significant clusters of adjacent data points (defined temporally and spatially). 

The results from the permutation test with the alternative hypothesis ${\mathit {confusing}>\mathit{clean}}$ reveal one significant cluster ($p = 0.031$) in the time interval from $390$ to $660~\text{ms}$ (cf. Figure~\ref{fig:frp}c). Although almost all electrodes are considered to be part of the significant cluster (at least, at one time point), the largest temporal extension of significant amplitude differences can be observed at midline frontal and fronto-central recording sites (i.e., Fz, FCz). Figure~\ref{fig:frp}c displays the significant cluster across time and electrodes (surrounding black lines) over the amplitude difference in the grand average. 

These results show that the late frontal positivity is significantly greater for confusing code than for clean code, indicating that both conditions are processed differently starting at around $390~\text{ms}$ after the onset of the first fixation.

\section{Discussion}\label{sec:discussion}

The primary goal of this study was to apply an FRP approach -- an established technique from psycholinguistics -- to achieve a better understanding of the neurocognitive processes that occur when reading confusing program code (i.e., code that contains atoms of confusion) by exploring parallels in terms of ERP components between the processing of program code and natural language processing in an ecologically valid experimental setting. Therefore, we investigated whether atoms of confusion trigger electrophysiological responses, which turned out to be similar to those engaged by unexpected but plausible inputs during language comprehension. In particular, our FRP analysis revealed a \emph{late positivity with a frontal scalp topography}. It onsets around $400~\text{ms}$ after the first eye fixation in the area containing the atom of confusion and was significantly larger for confusing code compared to clean code snippets. Notably, this component resembles other late frontal positivities found in natural language processing studies, in which the late frontal positivity is elicited by unexpected words, often but not necessarily in highly constraining contexts~\cite{federmeier2007thinking, holtje2022benefits, hubbard2024impact}. As outlined in Section~\ref{sec:introduction}, the late frontal positivity can be dissociated from the P600, a late ERP component, that is elicited by words that are perceived as highly implausible during online comprehension~\cite{Kuperberg2020, friederici2002towards}.


So what does the late frontal positivity reveal about the processes involved in processing an atom of confusion? In the literature on natural language comprehension, the predominant account for a late frontal positivity holds that it reflects expectancy mismatch related processing~\cite{federmeier2007thinking, holtje2022benefits} or late processing \emph{costs} due to the perceived difference between a prediction and the bottom-up input in highly constraining sentences~\cite{kutas1993company, ness2018lexical}. However, a range of studies has also observed that unexpected words within weakly constraining sentences can elicit a late frontal positivity~\cite{chow2016bag, freunberger2016semantic, Thornhill2012, zirnstein2018cognitive, ng2017use, hubbard2019downstream}.
Brothers et al.~\cite{brothers2020going, brothers2023multiple} proposed that the essential properties of a stimulus that lead to a late frontal positivity might be the presence of an unexpected but plausible word, which causes a substantial update of the situation model. More specifically, Kuperberg et al.~\cite{kuperbergP600review} suggest that late frontal positivities arise in situations where there is initial uncertainty about the situation model, which triggers the retrieval of information from long-term memory, enabling the comprehender to accumulate confidence in their ability to successfully update their situation model. In the case of high contextual constraint and unexpected lexical material, this occurs because the unexpected word first reduces confidence in the situation model, triggers retrieval from long-term memory, and finally leads to a high confidence decision that the situation model can be successfully updated~\cite{kuperbergP600review}. In the case of a low-constraint context and unexpected but plausible lexical input, retrieval of relevant event-level information from long-term memory allows the comprehender to accumulate confidence about how to update their situation model. Kuperberg et al.~\cite{kuperbergP600review} point out that the late frontal positivity effect in language processing seems to depend on the task, how engaged an individual is in deeply understanding the situation, and whether they can achieve confidence about the situation model based on the information they retrieved from long-term memory. 

These ideas align well with a plausible mechanism for processing the atoms of confusion: Atoms of confusion are infrequent in code~\cite{Gopstein2020} and hence generally unexpected, but they are neither implausible nor incorrect, as they reflect syntactically correct code patterns that can be unambiguously executed by the computer. However, when a comprehender encounters an atom of confusion, they are likely to be unsure at first about its effect on program execution and the variables they need to keep track of for the task (i.e., they are unsure about how to update their situation model), which is in line with observations that relate confusing code patterns to increased cognitive load~\cite{Yeh2017}. As a result, comprehenders access their long-term memory to remember exactly how to interpret the atom of confusion. If this retrieval is successful, the comprehender can confidently update their situation model. In the case of code without an atom, the situation model will also be updated, but there is no point of uncertainty followed by confidence on how to do that. This account also generates new predictions to be tested in future studies: We would expect that the late frontal positivity should only be present in trials where the comprehender could successfully resolve the ambiguity. It should therefore not appear in situations in which the atom of confusion contains a syntactically invalid code element, for which we would expect a P600 with a posterior scalp topography~\cite{Kuo2024}. Also, we do not expect a late frontal positivity to appear when the comprehender remains uncertain about the semantics of the atom or when they are given a different task that does not require updating the situation model. 

In summary, we presented, to our knowledge, the first-ever study that employs an FRP approach to explore the processing of confusing elements in program code. The results suggest that the brain engages highly similar mechanisms for processing unexpected but plausible elements in program code and natural language comprehension. This implies that studies on program comprehension can rely on established ERP components in natural language processing. The FRP approach offers new research perspectives with high ecological validity, focusing on the temporal characteristics of online program code comprehension. These results are significant for software engineering research, as they provide a foundation for a better understanding of code comprehension. Based on this common methodological ground, methods and findings from program comprehension research may be transferred back to research on natural language comprehension. The findings of our study, in particular, and this line of research, in general, also have far-reaching implications for the practice of software engineering. By shedding light on how programmers' brains process and respond to confusing code, we can enable programmers to develop and maintain more secure and reliable software. Specifically, our line of research supports programming practice by providing empirical evidence for coding conventions that avoid confusion-inducing code patterns. Moreover, programming language designers can leverage these insights to design consistent language constructs that minimize the potential for confusion and resulting severe mistakes. Finally, educators can incorporate these findings into their teaching practices, ultimately educating future programmers to apply more effective software development practices.

\section{Methods}\label{sec:methods}
\subsection{Participants}
To estimate the number of participants required to find the hypothesized effect, we performed a power analysis with $\mathit{power} = 0.8$, $\alpha = 0.05$, and $\mathit{dz} = 0.701$ (based on the reported $F(1,23) = 11.8$) for the late frontal positivity detected by Thornhill and Peterson~\cite{Thornhill2012} for high-constraint unrelated sentence endings. This analysis estimated a minimum of $18$ participants to correctly measure an existing effect (two-sided t-test). Due to various uncertainties in our setup and for the counterbalancing of the three version blocks across participants, we increased our sample to $24$ participants. While this number might appear low, it is in line with ERP research in general~\cite{Boudewyn2018ERPTrials,Clayson2019ERPSampleSizes,holtje2022benefits,hutzler2007welcome} and specifically with existing studies in the domain of software engineering~\cite{Kitchenham2024Samplesize,Peitek2021ProgramCA}. Further studies using larger samples shall be performed to generalize our results to a larger population, with an explicit focus on expert developers.

We recruited participants at our host institution (Saarland University) via e-mail lists, flyers, and (online) bulletin boards. We required participants to have at least an undergraduate degree in computer science or a related field. We further asked for basic experience with an object-oriented programming language (e.g., Java or C++), as the selected snippets use minimal language-specific operations. We only invited right-handed participants.

Our study was approved by the ethics committee of the Faculty of Human and Business Sciences at Saarland University. The experiment was conducted according to the relevant guidelines and regulations provided by the ethical review board. All participants provided their informed consent.

Overall, we invited $26$ participants to our experiment. To ensure high data quality, we excluded two participants from our analysis due to hardware malfunctions during the experiment or frequent excessive artifacts in the EEG data (see Section~\ref{sec:eeg_recording} and~\ref{sec:eeg_prepocessing}). $9$ out of the $24$ included participants were female, and $15$ were male, with a total age range from $21$ to $45$ years, and $14$ participants being younger than $26$. In Table~\ref{tab:participants}, we provide an overview of our included participants' demographics and their programming experience.

\begin{table}[ht]
\caption{Overview of the programming experience and activity of our participants}
\label{tab:participants}%
\centering
\begin{tabular*}{0.8\textwidth}{rllrp{0.3cm}llr}
\cmidrule[\heavyrulewidth](lr){1-4}\cmidrule[\heavyrulewidth](lr){6-8}
\multicolumn{2}{l}{Programming Experience} &\# &\multicolumn{1}{l}{\%} &\multirow{1}{3em}& Frequency of programming activity&\#& \multicolumn{1}{l}{\%} \\
\cmidrule(lr){1-4}\cmidrule(lr){6-8}
$<$1& Year&1&4.1&&Multiple times a year&2&8.3\\
1--3 &Years&7&29.2&&Multiple times a month&5&20.8\\
3--5 &Years&7&29.2&&1--2 times a week&4&16.6\\
5--10 &Years&8&33.3&&3--5 times a week&9&37.5\\
$>$10 &Years&1&4.1&&Every day&4&16.6\\
\cmidrule[\heavyrulewidth](lr){1-4}\cmidrule[\heavyrulewidth](lr){6-8}
\end{tabular*}
\end{table}

\subsection{Experimental variables}

The independent variable of our experiment is the snippet type, which has two conditions: either \emph{confusing} or \emph{clean}. As dependent variables, we measured the comprehension time, answer correctness, subjective difficulty rating, eye movements via an eye tracker, and brain activity via an EEG system for each snippet (see Figure~\ref{fig:design}).

\subsection{Stimuli}
The code snippets in our study are based on the snippet pairs from the study by Langhout and Aniche~\cite{Langhout2021}, which we adjusted to reduce fixations and confounding neurocognitive processes in the FRP analysis. In a pilot study, we rated these snippet pairs considering three criteria regarding the snippet structure, the area of the atom of confusion, and behavioral response (cf. supplementary information, Appendix A). Specifically, we required that within each snippet pair, the snippets have a comparable structure to reduce the influence of confounding factors. This was determined by the number and length of the code lines. Furthermore, to facilitate the selection of suitable fixations for FRP analysis, each snippet in a pair must contain a single clear area of interest that either includes the atom of confusion (for confusing snippets) or a clarified equivalent (for clean snippets). Finally, we tested whether the snippets were of similar complexity and whether they were, in principle, understandable to students of our target population (familiarity with the syntax of Java) in a pilot study. To this end, we set a threshold of $50\%$ answer correctness or $50\%$ subjective difficulty rating as easy for either condition. Snippets that showed a clear violation of the expected behavior by lower correctness or fewer easy ratings were excluded from the stimulus materials.

We used $144$ snippets ($72$ snippet pairs), each pair consisting of one confusing and one clean snippet, which are semantically equivalent and differ only in the presence of an atom of confusion. We presented only one condition of each snippet pair to each participant ($72$ snippets), thereby avoiding learning effects. In total, each participant viewed an equal number of snippets for both conditions ($36$ snippets each). 

While the combined eye-tracking and EEG approach offers a higher ecological validity than using a token-wise presentation which is common in ERP studies with linguistic events, it is important to acknowledge that constraints inherent to eye tracking and EEG remain. These constraints limit the length and, consequently, the ecological validity of the presented code snippets. Nevertheless, because atoms of confusion consist of small, localized patterns, they can be meaningfully embedded within code even under these length limitations.

\subsection{Procedure}
At the beginning of the experiment, all participants provided their informed consent and completed a set of questionnaires, including a basic demographic questionnaire~\cite{Zhuang2023, Peitek2020}. To assess programming experience, we inquired about the years and recent frequency of programming activities, as well as all programming languages encountered and recently used. This provides us with a broad estimate of their capabilities and background~\cite{Siegmund2014}. Additionally, participants filled out the Edinburgh Handedness Inventory~\cite{Oldfield1971} to ensure right-handedness. 

After setting up the EEG (see Section~\ref{sec:eeg_recording}), the participants were seated in front of the eye tracker attached to the screen. Then, we explained the purpose and setup of the study with introductory slides. To familiarize the participants with the task, they performed a brief training session consisting of one trial.

The experiment was divided into three consecutive blocks with $24$ trials each. Between each block, participants took a break of ten minutes to reduce learning and fatigue effects. We pseudo-randomized the presentation order with the constraints that no more than three snippets of one condition appeared in a row and that at least two other snippets appeared between two snippets containing the same atom of confusion. Each trial started with the presentation of a fixation cross for $5~\text{s}$. Next, the code snippet was presented (see Figure~\ref{fig:design}). The participants' task was to determine the output of variable \verb|R| (as in \textbf{Result}) in each snippet as quickly and accurately as possible. Participants had up to $30~\text{s}$ to comprehend each snippet, but they could proceed earlier after at least $3~\text{s}$ by hitting the \emph{enter} key. Then, we presented four response options: One option was the correct value, two were incorrect values that would follow typical erroneous thought processes associated with the respective atom (only one for two-choice answer options), and one \emph{I don't know} option. The last option was intended to prevent participants from guessing. Finally, we asked participants whether the code snippet was easy or difficult to comprehend, also allowing the \emph{I don't know} option. 

After the experiment, we conducted a semi-structured interview that included questions on the participants' views about the experiment and each snippet. Finally, we paid the compensation of $€12$ per hour, calculated on a quarter-hourly basis.

\subsection{Recording and preprocessing}

\subsubsection{Behavioral data preprocessing} 
After extracting the behavioral data, we checked whether each participant was generally able to perform the tasks based on their overall answer correctness, which we compared to the estimated correctness based on chance (i.e., $38\%$ due to $21$ snippets providing two and $51$ snippets providing three answer options). All participants achieved a correctness of at least $55\%$ ($79.33\%~\pm 8.46\%$). Thus, no participants were excluded on the basis of task performance. One trial had to be excluded from the behavioral analysis due to a hardware malfunction, leaving us with $1,727$ trials for the behavioral analysis. We transformed the subjective difficulty rating into a binary variable ($1$ if \emph{easy}, $0$ for \emph{hard}, \emph{I don't know} , or no answer), since we can assume that every participant who did not immediately answer the question with ease has experienced at least some difficulty.

\subsubsection{Eye-movement recording} 
We recorded the eye movements using a Tobii Pro Spectrum eye tracker with a sampling rate of $1,200~\text{Hz}$ and positioned participants using the Tobii Eye Tracker Manager to ensure an optimal head position. Each block began with calibration and validation, checking all validation points to ensure high accuracy and precision. If there was a clear offset in the validation points, if some points did not receive any data, or if the data for a point were scattered across the screen, we repositioned the participant and restarted with calibration and validation.

\subsubsection{Eye-movement preprocessing} 
The eye-tracking data, specifically the fixations, are the basis for our FRP analysis to reliably detect potential confusion at the correct moment. Missing a fixation or selecting an incorrect fixation would substantially affect the quality of our FRP analysis. Therefore, only trials with high data quality, including data correction, were considered in our eye-movement analysis.

To this end, we separated the raw eye-tracking data of all $1,727$ trials based on the start and end of the snippet presentations. While the quality was overall high, for some trials, the eye tracker failed to identify the gaze position (i.e., resulting in missing values in the data). Because this poses a risk of missing a relevant fixation in the area of interest, we first excluded trials in which more than $25\%$ of the snippet presentation consisted of missing values. We also removed trials that contained one or more intervals of missing values exceeding one second. Based on this, $34$ trials were excluded, and $1,693$ trials were left for further analyses.

As a next step, to reliably detect the correct fixation in the area of the atom of confusion, we manually checked the quality of all fixations. To this end, we first detected fixations and saccades in the raw eye-tracking data with I2MC, which is a noise-robust algorithm~\cite{Hessels2017} with a minimal accepted fixation duration of $200~\text{ms}$. Then, we manually reviewed each trial for fixations that are clear outliers (i.e., single fixations that could not be associated with a code line). $202$ trials (out of $1,693$) clearly contained outlier fixations. For these trials, we eliminated the outlier fixations ($344$) from the fixation data.

Another risk to data quality arises from fixations with an offset on the y-axis, because this can lead to assigning fixations to incorrect code lines. To reduce the risk, we used two automated fixation correction algorithms described by Carr~et~al.~\cite{Carr2022}, which are suitable for code reading (i.e., no assumption of linear line-by-line reading), namely \emph{cluster} and \emph{stretch}. We fine-tuned these algorithms, because they both tend to place fixations in as many code lines as possible, although participants may ignore code lines that do not hold information relevant for program execution and variable values (i.e., lines containing only brackets \verb|{| or \verb|}|). Specifically, we derived variants that ignore such code lines in favor of other line matches. For each trial, we generated plots to compare the results of the four fixation corrections against the original fixation positioning and, for each fixation correction, to assess the precision on the code lines and suitability for detecting fixations within or near the area of interest later on. Two members of the research team manually reviewed the automated fixation corrections independently of each other. Upon a disagreement ($275$ out of $1,693$ trials), the differences were discussed in detail, on $34$ occasions with a third team member to reach consensus, resulting in an almost perfect inter-rater reliability of $81\%$ determined by Cohen's Kappa~\cite{Landis1977}. During these reviews, we detected that, for a few participants, there was also a noticeable offset on the x-axis. Since this offset was usually constant within a block, we corrected it with a constant offset.

We excluded trials from further analysis when we were unable to confidently align the fixations to the code ($192$ out of $1,693$). This was the case when fixations were either entirely off the visualized code and could not be mapped to their correct position, or when single fixations could not be confidently assigned to the correct line. A visualization of all original fixations contrasted with our correction is part of the replication package.

For each of the remaining $1,501$ trials, we selected the first fixation within the area of interest as the time point at which the atom of confusion was processed. $68$ trials for which we did not receive a result using this strategy (i.e., the area of interest may not be fixated) were excluded from the FRP analysis. In total, we excluded $294$ trials in the eye-movement preprocessing, which resulted in $1,433$ trials usable for the FRP analysis.

\subsubsection{EEG recording}
\label{sec:eeg_recording}
We recorded continuous EEG from $28$ scalp electrodes (Fp1, Fp2, F7, F3, Fz, F4, F8, FC5, FC3, FCz, FC4, FC6, T7, C3, Cz, C4, T8, CP3, CPz, CP4, P7, P3, Pz, P4, P8, O1, O2, and A2) placed in a cap according to the International 10--20 system~\cite{Jasper1958} using BrainVision Recorder (Brain Products). Electrode AFz served as the ground electrode. An additional four electrodes were placed to the left of the left eye as well as above, beneath, and to the right of the right eye to monitor eye movements and blinks (electrooculography; EOG). Two more electrodes were placed over the left and right mastoids. Electrodes were referenced to the electrode placed over the left mastoid during recording. EEG signals were amplified with a bandpass filter of $0.016$ to $250~\text{Hz}$ using a 16 bit BrainAmp DC amplifier. We treated all electrode locations with gel and cotton swabs to achieve an impedance below $5~\text{k}\Omega$ for EEG and below $10~\text{k}\Omega$ for EOG electrodes. Continuous analog-to-digital conversion of the EEG and stimulus trigger codes was performed at a sampling frequency of $500~\text{Hz}$.

\subsubsection{EEG preprocessing}
\label{sec:eeg_prepocessing}
We manually marked excessive artifacts caused by high tension or muscle movement, such as clenching teeth, using BrainVision Analyzer. Then, we applied two finite-impulse-response filters, consisting of a bandpass filter of $0.05$ to $30~\text{Hz}$ of order 4 and a notch filter at $50~\text{Hz}$. Next, we employed an independent component analysis (ICA) using the classical biased restricted info-max algorithm implemented in Brain Vision Analyzer 2.1 on the filtered continuous EEG data (excluding marked segments) to extract components of the signal that are associated with artifacts caused by eye or muscle artifacts. Next, in the Inverse ICA, we set the weight of these components on the final signal to $0$. As a last step, we re-referenced the EEG channels to the mean of both mastoid electrodes to avoid bias from using one hemisphere as the reference.

After this preparation, we synchronized the event triggers in the EEG files with the behavioral data. For two trials in the behavioral data, we were unable to assign the corresponding EEG annotations due to hardware malfunctions. Then, we cut each participant's continuous EEG file into individual trials (from the start of the fixation cross to the end of the snippet presentation), resulting in $1,725$ trials usable for ERP analysis. For the FRP analysis, this leaves us with $1,432$ ($=1,725-294+1$) trials due to an overlap of one trial in the excluded trials within the EEG and the eye-tracking data preparation. 

In summary, we obtained $1,727$ trials fulfilling the requirements for the behavioral data analysis. Among these, $1,433$ trials were suitable for eye-movement analysis, and $1,725$ were suitable for ERP analysis. Merging these two sets left us with an overlap of $1,432$ trials to be used in the FRP analysis.

\subsection{Data analysis}
\subsubsection{Behavioral analysis}
\label{sec:behavioral_analysis}
For each of the $1,727$ trials resulting from the processing of only the behavioral data, we obtained measures of comprehension time, answer correctness, and the subjective difficulty rating.

We analyzed the behavioral data using linear mixed effects models. As a model family, we chose a Gaussian error model with logarithmic values for fixation duration. Answer correctness and subjective difficulty ratings were coded as binary variables ($\mathit{correct} = 1$, $\mathit{incorrect} = 0$ and $\mathit{easy} = 1$, $\mathit{other} = 0$, respectively) and analyzed using binomial mixed effects models. For each behavioral measure, the regression models included the fixed parameter condition to identify whether a trial was confusing or clean (Condition: $\mathit{confusing} = 1$, $\mathit{clean} = 0$), as well as the block number (BlockNo: $1$--$3$) and the trial number within a block (ItemOrder: $1$--$24$) along with their interactions to account for potential learning and habituation effects. Linear mixed effects models included random intercepts and random slopes for condition under participants and items; the fixed effects structure was reduced using backward model selection based on the Akaike information criterion and a significance threshold of $p<0.05$.

\subsubsection{FRP analysis}
To calculate the FRPs, we truncated the EEG segment of each trial relative to the start of the fixation into the epochs from $-300$ to $1000~\text{ms}$ including a baseline from  $-300$ to $0~\text{ms}$. The offset of the fixation compared to stimulus onset leads to a further reduction of trials since fixations at the end of the trial might not have the entire epoch length available within the snippet view and thus cannot be included, leaving us with $1405$ ($= 1432 - 27$) trials. Next, the epochs were baseline-corrected to reduce the influence of pre-stimulus activity. That is, for each electrode and each sample point within the epoch ($-300$ to $1000~\text{ms}$), the average pre-stimulus voltage was subtracted from the EEG signal. For determining trials containing artifacts, we ignored all EOG electrodes as well as the reference electrodes. We excluded all epochs that violated one of the following voltage constraints on any of the electrodes, which indicates the presence of an artifact in the epoch: First, a voltage step between two consecutive samples must be lower than $30~$\textmu$\text{V}$. Second, the maximal voltage difference within $0.2~\text{s}$ should not exceed $100~$\textmu$\text{V}$. Last, the absolute amplitude after baseline correction must remain below$~\pm 70~$\textmu$\text{V}$ at any point in time within the epoch. We excluded $18$ trials ($1387 = 1405 - 18$ remaining). As a result, on average, $29$ ($\mathit{SD}=5$; range: $16$--$35$) trials per condition and participant are suitable for the FRP analysis. Finally, we averaged the epochs per participant, condition, and electrode. 

For statistical analysis, we calculated the FRP (${\mathit {confusing}-\mathit{clean}}$) difference wave per participant and electrode to use it as a dependent variable in a pairwise cluster permutation test~\cite{candia2022cluster}. This test identifies possibly significant data points consisting of time points and electrodes in which the test statistics exceed the significance threshold ($p<0.05$) of a single paired t-test, and clusters these if at least two spatially or temporally adjacent points reach the significance threshold. Spatial adjacency was defined by adjacent electrodes (e.g., Fp1 and Fp2, see Figure~\ref{fig:frp}a) and temporal adjacency was defined by two consecutive time points. Then, the test evaluates the significance of each detected cluster by a permutation test, which has the advantage of inherently controlling for multiple testing~\cite{Groppe2011}. To maximize the power in the permutation test, we followed the suggestions of Luck et al.~\cite{Luck2014}: The epochs were down-sampled to 100 Hz, and electrodes T7, T8, O1, and O2 were not included in the analysis since we did not expect any significant effects at these electrode sites.

Additionally, we reduced the analysis interval to $100$ to $1,000~\text{ms}$. We performed two pairwise cluster permutation tests with this input for the difference being both a significantly smaller (confusing<clean) and a significantly higher (confusing>clean) amplitude than $0~$\textmu$\text{V}$ and considered all resulting clusters with an alpha level of $ p < 0.05 $ as significant.
For each of the permutation tests, we performed the maximum number of permutations ($16,777,216$)~\cite{Groppe2011}.

\section*{Acknowledgements}
We would like to thank all the participants of the study who made this research possible, as well as our research assistants, Youssef Abdelsalam, Yannick Lehmen, and Marcellius William Suntoro, who helped with the technical setup.

\section*{Funding}
This work has been supported by the European Union as part of the ERC Advanced Grant \emph{Brains On Code} (101052182) and by the German Research Foundation as part of the Transregional Collaborative Research Center 248 \emph{Foundations of Perspicuous Software Systems} (389792660). Axel Mecklinger and Regine Bader were funded by the German Research Foundation as part of the Collaborative Research Center 1102 Information Density and Linguistic Encoding (232722074), project A6. 

\section*{Competing interests}
There are no competing interests.


\section*{Data availability}
We provide all relevant data in line with open data principles under a CC-BY license, respecting our local privacy laws (GDPR). 
Specifically, the datasets collected during the experiment and generated for the analysis are long-term archived in the Zenodo repository \textit{Dataset for  ``Fixation-related potentials reveal that confusing program code elicits a late frontal positivity''}, \href{https://doi.org/10.5281/zenodo.14229848}{https://doi.org/10.5281/zenodo.14229848}.

\section*{Materials availability}
\label{subsec:materials-availability}
This experiment used code available in Python (v3.11.5) as well as R (v4.3.2) for linear mixed effects models. Additionally, we used other open-source (PsychoPy, v2021.2.3) and commercial applications (Tobii Eye-Tracker Manager (v2.6.0), BrainVision Recorder (v1.20.0801), and BrainVision Analyzer (v2.3.0.8300)) to perform the experiment. We deposited the scripts and content in a GitHub Repository \href{https://github.com/brains-on-code/AoC-FRP-Code}{https://github.com/brains-on-code/AoC-FRP-Code} and they are available for unrestricted open access under a CC-BY license.

\bibliography{References}

\section*{Author contribution}
Annabelle Bergum, Anna-Maria Maurer, Norman Peitek, Regine Bader, Axel Mecklinger, Janet Siegmund, and Sven Apel developed the concept and the general research idea.
Annabelle Bergum, Anna-Maria Maurer, Norman Peitek, Regine Bader, Axel Mecklinger, and Sven Apel devised the study design and formulated the method.
Anna-Maria Maurer conducted the study.
Anna-Maria Maurer, Norman Peitek, and Vera Demberg performed the data analysis.
All authors contributed to the interpretation, writing, editing, and reviewing.
Anna-Maria Maurer and Annabelle Bergum prepared the visualizations.
Axel Mecklinger, Regine Bader, Sven Apel, Vera Demberg, and Janet Siegmund provided the funding and supervised the project.


\end{document}